\documentclass{emulateapj}
\usepackage{apjfonts}

\newcommand{\pivec}{\mbox{\boldmath $\pi$}}

\begin{document}

\title{OGLE-2005-BLG-153: Microlensing Discovery and Characterization of A
Very Low Mass Binary}

\author{
K.-H. Hwang$^{1}$,
A.\ Udalski$^{2}$,
C. Han$^{1,39}$,
Y.-H. Ryu$^{1}$,
I.A.\ Bond$^{4}$,
J.-P. Beaulieu$^{5}$,
M.\ Dominik$^{6}$,
K. Horne$^{6}$,
A.\ Gould$^{7}$,
B.S. Gaudi$^{7}$\\
and\\
M. Kubiak$^{2}$, 
M.K. Szyma\'nski$^{2}$,
G.\ Pietrzy\'nski$^{2,8}$, 
I. Soszy\'nski$^{2}$,
O. Szewczyk$^{2,8}$, 
K.\ Ulaczyk$^{2}$, 
{\L}. Wyrzykowski$^{2,9}$ \\
(The OGLE Collaboration), \\
F. Abe$^{10}$,
C.S.\ Botzler$^{12}$,
J.B.\ Hearnshaw$^{13}$,
Y.\ Itow$^{10}$,
K.\ Kamiya$^{10}$,
P.M.\ Kilmartin$^{14}$,
K. Masuda$^{10}$,
Y. Matsubara$^{10}$,
M. Motomura$^{10}$,
Y. Muraki$^{16}$,
S. Nakamura$^{10}$,
K. Ohnishi$^{17}$,
C. Okada$^{10}$,
N. Rattenbury$^{12}$,
To. Saito$^{18}$,
T. Sako$^{10}$,
M. Sasaki$^{10}$,
D.J. Sullivan$^{15}$,
T.\ Sumi$^{10}$,
P.J. Tristram$^{15}$, 
J.N. Wood$^{12}$,
P.C.M. Yock$^{12}$,
T. Yoshioka$^{10}$\\
(The MOA Collaboration), \\
  M. Albrow$^{26}$,
  D.P.\ Bennett$^{11}$,
  D.M.\ Bramich$^{19}$,
  S. Brillant$^{24}$,
  J.A.R. Caldwell$^{35}$,
  J.J. Calitz$^{36}$,
  A. Cassan$^{5}$,
  K.H. Cook$^{37}$,
  E. Corrales$^{5}$,
  C. Coutures$^{20}$,
  M. Desort$^{5}$,
  S. Dieters$^{22}$,
  D. Dominis$^{21}$,
  J. Donatowicz$^{19}$
  P. Fouqu\'e$^{27}$,
  J. Greenhill$^{22}$,
  K. Harps{\o}e$^{23,38}$,
  K. Hill$^{22}$,
  M. Hoffman$^{21}$,
  U.G. J\o rgensen$^{23,38}$,
  S. Kane$^{28}$,
  D. Kubas$^{5}$,
  R. Martin$^{25}$,
  J.-B. Marquette$^{5}$,
  P. Meintjes$^{36}$,
  J. Menzies$^{29}$,
  K. Pollard$^{26}$,
  K. Sahu$^{30}$,
  I. Steele$^{32}$,
  C. Vinter$^{23}$,
  J. Wambsganss$^{31}$,
  A. Williams$^{25}$,
  K. Woller$^{23}$,
  M. Burgdorf$^{32}$,
  C. Snodgrass$^{33}$,
  M. Bode$^{32}$\\
(The Planet/RoboNet Collaboration);\\
and\\
D.L. Depoy$^{34}$,
C.-U. Lee$^{3}$,
B.-G. Park$^{3}$,
R.W. Pogge$^{7}$
(The $\mu$FUN Collaboration)
}

\begin{abstract}
The mass function and statistics of binaries provide important 
diagnostics of the star formation process. Despite this importance, 
the mass function at low masses remains poorly known due to 
observational difficulties caused by the faintness of the objects. 
Here we report the microlensing discovery and characterization 
of a binary lens composed of very low-mass stars just above the 
hydrogen-burning limit. From the combined measurements of the 
Einstein radius and microlens parallax, we measure the masses of the
binary components of $0.10\pm 0.01\ M_\odot$ and $0.09\pm 0.01\
M_\odot$. This discovery demonstrates that microlensing will 
provide a method to measure the mass function of all Galactic 
populations of very low mass binaries that is independent of 
the biases caused by the luminosity of the population.
\end{abstract}

\section{Introduction}

\footnote{
{Department of Physics, Chungbuk National University, Cheongju 361-763, Republic of Korea.}\\
$^{2}${Warsaw University Observatory, Al. Ujazdowskie 4, 00-478 Warszawa, Poland.}\\
$^{3}${Korea Astronomy and Space Science Institute, Daejeon 305-348, Republic of Korea.}\\
$^{4}${Institute of Information and Mathematical Sciences, Massey University, Private Bag 102-904, North Shore Mail Centre, Auckland, New Zealand.}\\
$^{5}${Institut d'Astrophysique de Paris, CNRS, Universit\'e Pierre et Marie Curie UMR7095, 98bis Boulevard Arago, 75014 Paris, France.}\\
$^{6}${Scottish Universities Physics Alliance, University of St Andrews, School of Physics and Astronomy, North Haugh, St Andrews KY16 9SS, UK.}\\
$^{7}${Department of Astronomy, The Ohio State University, 140 W. 18th Ave., Columbus, OH 43210, USA.}\\
$^{8}${Universidad de Concepci\'on, Departamento de Fisica, Casilla 160-C, Concepci\'on, Chile.}\\
$^{9}${Institute of Astronomy, University of Cambridge, Madingley Road, Cambridge CB3 0HA, UK.}\\
$^{10}${Solar-Terrestrial Environment Laboratory, Nagoya University, Nagoya, 464-8601, Japan.}\\
$^{11}${Department of Physics, University of Notre Dame, Notre Dame, IN 46556, USA.}\\
$^{12}${Department of Physics, University of Auckland, Private Bag 92019, Auckland, New Zealand.}\\
$^{13}${University of Canterbury, Department of Physics and Astronomy, Private Bag 4800, Christchurch 8020, New Zealand.}\\
$^{14}${Mt. John Observatory, P.O. Box 56, Lake Tekapo 8770, New Zealand.}\\
$^{15}${School of Chemical and Physical Sciences, Victoria University, Wellington, New Zealand.}\\
$^{16}${Department of Physics, Konan University, Nishiokamoto 8-9-1, Kobe 658-8501, Japan.}\\
$^{17}${Nagano National College of Technology, Nagano 381-8550, Japan.}\\ 
$^{18}${Tokyo Metropolitan College of Industrial Technology, Tokyo 116-8523, Japan.}\\
$^{19}${European Southern Observatory, Karl-Schwarzschild-Stra$\beta$e 2, 85748 Garching bei M\"{u}nchen, Germany.}\\
$^{20}${CEA DAPNIA/SPP Saclay, 91191 Gif-sur-Yvette cedex, France.}\\
$^{21}${Universit\"at Potsdam, Institut f\"ur Physik, Am Neuen Palais 10, 14469 Potsdam, Astrophysikalisches Institut Potsdam, An der Sternwarte 16, D-14482, Potsdam, Germany.}\\
$^{22}${University of Tasmania, School of Mathematics and Physics, Private Bag 37, Hobart, TAS 7001, Australia.}\\
$^{23}${Niels Bohr Institute, University of Copenhagen, Juliane Maries Vej 30, 2100 Copenhagen, Denmark.}\\
$^{24}${European Southern Observatory, Casilla 19001, Santiago 19, Chile.}\\
$^{25}${Perth Observatory, Walnut Road, Bickley, Perth, WA 6076, Australia.}\\
$^{26}${University of Canterbury, Department of Physics and Astronomy, Private Bag 4800, Christchurch 8020, New Zealand.}\\
$^{27}${Observatoire Midi-Pyr\'en\'ees, Laboratoire d'Astrophysique, UMR 5572, Universit\'e Paul Sabatier-Toulouse 3, 14 avenue Edouard Belin, 31400 Toulouse, France.}\\
$^{28}${Department of Astronomy, University of Florida, 211 Bryant Space Science Center, Gainesville, Florida 32611-2055, USA.}\\
$^{29}${South African Astronomical Observatory, PO Box 9, Observatory 7935, South Africa.}\\
$^{30}${Space Telescope Science Institute, 3700 San Martin Drive, Baltimore, Maryland 21218, USA.}\\
$^{31}${Astronomisches Rechen-Institut (ARI), Zentrum f\"ur Astronomie, Universit\"at Heidelberg, M\"onchhofstrasse 12-14, 69120 Heidelberg, Germany.}\\
$^{32}${Astrophysics Research Institute, Liverpool John Moores University, Twelve Quays House, Egerton Wharf, Birkenhead CH41 1LD, UK.}\\
$^{33}${Max Planck Institute for Solar System Research, Max-Planck-Str. 2, 37191 Katlenburg-Lindau, Germany.}\\
$^{34}${Department of Physics, Texas A\&M University, College Station, TX, USA.}\\
$^{35}${McDonald Observatory, 16120 St Hwy Spur 78 \#2, Fort Davis, Texas 79734, USA}\\
$^{36}${Boyden Observatory, University of the Free State, Department of Physics, PO Box 339, Bloemfontein 9300, South Africa.}\\
$^{37}${Lawrence Livermore National Laboratory, IGPP, PO Box 808, Livermore, California 94551, USA}\\
$^{38}${Centre for Star and Planet Formation, Geological Museum, {\O}ster Voldgade 5-7, 1350 Copenhagen, Denmark.}\\
$^{39}${Corresponding author}\\
}

\section{Introduction}

Microlensing occurs when a foreground astronomical object (lens) is
closely aligned to a background star (source) and the light from the
source star is deflected by the gravity of the lens \citep{einstein36}.
The phenomenon causes splits and distortions of the source star image.
For source stars located in the Milky Way Galaxy, the separation between
the split images is of order milli-arcsecond and thus the individual 
images cannot be directly observed. However, the phenomenon can be 
photometrically observed through the brightness change of the source 
star caused by the change of the relative lens-source separation 
\citep{paczynski86}.  Since the first discovery in 1993 \citep{alcock93, 
udalski93}, there have been numerous detections of microlensing events 
toward the Large and Small Magellanic Clouds, M31, and mostly the 
Galactic bulge fields.  Currently, microlensing events are being 
detected at a rate of nearly 1000 events per year \citep{udalski08, 
bond02}.

The properties of multiple systems such as binary frequency and mass
function provide important constraints for star formation theories,
enabling a concrete, qualitative picture of the birth and evolution 
of stars. At very low masses down to and below the hydrogen burning 
minimum mass, however, our understanding of formation processes is 
not clear due to the difficulties of observing these objects.  Over 
the last decade, there have been several searches for very low mass 
binaries [see reviews by \citet{basri00, oppenheimer00, kirkpatrick05, 
burgasser07}]. Despite these efforts, the number of very low mass 
binaries\footnote{An updated list of binaries with total masses 
below $0.2\ M_\odot$ can be found at the ``Very-Low-Mass Binary 
Archive'' (http://ldwarf.ipac.caltech.edu/vlm/), and contains 99 
entries as of Jan. 2010.} is not big enough to strongly constrain
their formation processes.

Microlensing occurs regardless of the brightness of lens objects 
and thus it is potentially an effective method to investigate the 
mass function of low-mass binaries. For lensing events caused by 
single-mass objects, it is difficult to measure the lens mass 
because the Einstein time scale $t_{\rm E}$, which is the only 
observable that provides information about the lens for general 
lensing events, results from the combination of the mass and distance 
to the lens and the transverse speed between the lens and source. 
The degeneracy can be partially lifted by measuring either an 
Einstein radius or a lens parallax and can be completely broken 
by measuring both. Einstein radii are measured from the deviation 
in lensing light curves caused by the finite-source effect 
\citep{gould94}. Most microlensing events produced by binaries 
are identified from the anomalies involved with caustic approaches 
or crossings during which the finite-source effect is important
\citep{nemiroff94, witt94}.
Therefore, Einstein radii can be routinely measured for the majority 
of binary-lens events. The microlens parallax is defined by 
\begin{equation}
\pi_{\rm E}= {\pi_{\rm rel}\over \theta_{\rm E}},
\label{eq1}
\end{equation}
where $\pi_{\rm rel}={\rm AU}(D_{\rm l}^{-1}-D_{\rm S}^{-1})$ is 
the lens-source relative parallax and $D_{\rm L}$ and $D_{\rm S}$ 
are the distances to the lens and source, respectively.  In general, 
parallaxes can be measured for events that last long enough that 
the Earth's motion can no longer be approximated as rectilinear 
during the event \citep{gould92}. The chance to measure the lens 
parallax for binary-lens events is higher than that of single-lens 
events because the average mass of binaries is bigger and thus 
time scales tend to be longer. In addition, the well-resolved 
caustic-crossing part of lensing light curves provide strong 
constraints on the lensing parameters and thus helps to pin-down 
enough anchor points on the light curve to extract otherwise 
too-subtle parallax effects \citep{an01}.

The number of binary-lens events with well-resolved anomalies is
increasing with the advance of observational strategies such as 
the alert system and follow-up observations. The increase of the 
monitoring cadence of existing and planned survey experiments will 
make the detection rate even higher. Although binary microlensing 
is biased toward separations similar to the Einstein radius, it 
is easily quantifiable for next-generation experiments that have 
continuous ``blind'' monitoring. Therefore, microlensing will be 
able to provide an important method to discover very low mass 
binaries and to investigate their mass function.

In this paper, we present the microlensing discovery and characterization
of a very low mass binary. We use this discovery to demonstrate that
microlensing will provide a method to measure the mass function of very
low-mass binaries that is free from the biases and difficulties of
traditional methods.

\begin{figure*}[t]
\epsscale{0.8}
\plotone{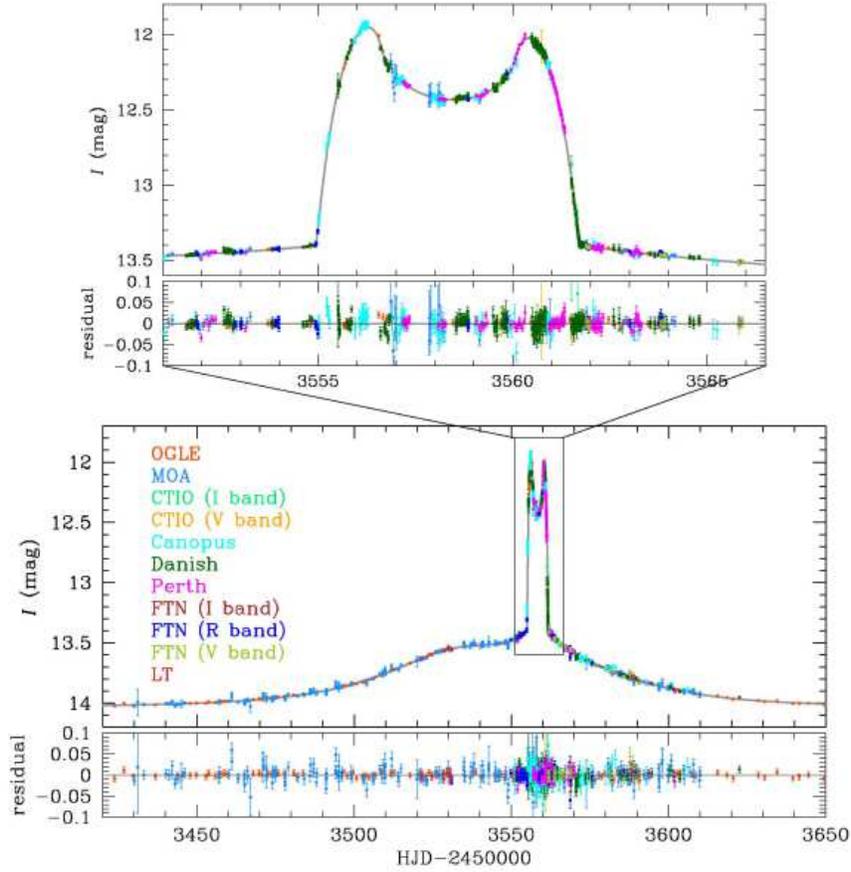}
\caption{\label{fig:one}
Light curve of the microlensing event OGLE-2005-BLG-153.  Also 
presented is the model curve for the best-fit solution. The upper
panel shows a zoom of the caustic-induced perturbation region.  
}\end{figure*}

\section{Observations}

The low-mass binary was detected from observations of the microlensing 
event OGLE-2005-BLG-153.  The event occurred on a Galactic bulge star 
located at right ascension $\alpha=18^{\rm h}04^{\rm m} 17^{\rm s}
\hskip-2pt .30$ and declination $\delta=-28^\circ 40'49''\hskip-2pt .3$ 
(J2000), which corresponds to the Galactic coordinates 
$(l,b)=(2.3^\circ, -3.4^\circ)$.  The event was detected by the Optical 
Gravitational Lensing Experiment (OGLE) using the 1.3 m Warsaw telescope 
of Las Campanas Observatory in Chile and was announced as a probable 
microlensing event on 14 April 2005.  The event was independently 
identified by the Microlensing Observation in Astrophysics (MOA) as 
MOA-2005-BLG-023 using the 0.6 m of Mt. John Observatory in New Zealand.

An anomaly alert was issued on 3 July 2005 by the Probing Lensing 
Anomalies Network (PLANET) collaboration.  Following the alert, 
the PLANET, RoboNet, and Microlensing Follow-Up Network ($\mu$FUN) 
teams intensively observed the event by using eight telescopes 
located on three different continents.  These telescopes include 
the PLANET 1.54 m Danish Telescope of La Silla Observatory in Chile, 
PLANET 1.0 m of Mt.\ Canopus Observatory in Australia, PLANET 0.6 m 
of Perth Observatory in Australia, $\mu$FUN 1.3 m SMARTS telescope 
of CTIO in Chile, RoboNet 2.0 m Faulkes Telescope S. (FTS) in Australia, 
RoboNet 2.0 m Faulkes Telescope N.  (FTN) in Hawaii, and RoboNet 2.0 m 
Liverpool Telescope (LT) in La Palma, Spain. Thanks to the follow-up 
observations, the light curve was densely resolved.

The event was analyzed before by \cite{skowron07} based on the data 
from OGLE observations as one of 9 binary-lens events detected in 
2005 season.  Here we reanalyze the event in depth with the addition 
of data from follow-up observations focusing on the physical parameters 
of the lens system.

\section{Modeling}

Figure 1 shows the light curve of the event.  It is characterized by 
the sharp rise and fall occurring at around the heliocentric Julian 
dates (HJD) of 2453556 and 2453560.  These features are caused by the 
crossings of the source star across a caustic, which represents a 
set of source positions at which the lensing magnification of a point 
source becomes infinite.  Therefore, the existence of such a feature 
immediately reveals that the lens is composed of two masses \citep{mao91}.

\begin{deluxetable*}{ccccccc}[h]
\tablecaption{Fit Parameters\label{table:one}}
\tablewidth{0pt}
\tablehead{
\multicolumn{1}{c}{} &
\multicolumn{1}{c}{standard} & 
\multicolumn{2}{c}{parallax} & 
\multicolumn{2}{c}{parallax+orbit} \\
\multicolumn{1}{c}{} &
\multicolumn{1}{c}{} &
\multicolumn{1}{c}{$(u_0>0)$} &
\multicolumn{1}{c}{$(u_0<0)$} &
\multicolumn{1}{c}{$(u_0>0)$} &
\multicolumn{1}{c}{$(u_0<0)$} 
}
\startdata
$\chi^2$             & 2406.812              & 1836.199              & 1982.545              & {\bf 1672.572}              & 1674.966                \\
$t_0$ (HJD')  & 3549.8286$\pm$0.0556  & 3551.2263$\pm$0.0950  & 3549.4179$\pm$0.0692  & {\bf 3549.1650$\pm$0.0992}  & 3548.6344$\pm$0.1452    \\
$u_0$                &    0.502$\pm$0.002    &    0.481$\pm$0.003    &   -0.515$\pm$0.002    &   {\bf 0.566$\pm$0.003}     &   -0.572$\pm$0.004      \\
$t_{\rm E}$ (days)   &   46.428$\pm$0.111    &   48.735$\pm$0.149    &   50.019$\pm$0.229    &   {\bf 44.541$\pm$0.121}    &   44.952$\pm$0.173      \\
$s$                  &    0.837$\pm$0.001    &    0.836$\pm$0.001    &    0.832$\pm$0.001    &   {\bf 0.848$\pm$0.001}     &    0.847$\pm$0.001      \\
$q$                  &    0.848$\pm$0.003    &    0.848$\pm$0.003    &    0.833$\pm$0.003    &   {\bf  0.871$\pm$0.003}    &    0.865$\pm$0.002      \\
$\alpha$             &    5.902$\pm$0.001    &    5.956$\pm$0.003    &    0.373$\pm$0.003    &   {\bf  5.900$\pm$0.002}    &    0.396$\pm$0.003      \\
$\rho_\star$         &    0.017$\pm$0.001    &    0.017$\pm$0.001    &    0.016$\pm$0.001    &   {\bf  0.018$\pm$0.001}    &    0.018$\pm$0.001      \\
$\Gamma_I$           &    0.571$\pm$0.019    &    0.510$\pm$0.016    &    0.545$\pm$0.016    &   {\bf  0.501$\pm$0.014}    &    0.495$\pm$0.013      \\
$\pi_{{\rm E},N}$    &                       &    0.762$\pm$0.040    &    0.555$\pm$0.078    &   {\bf  0.106$\pm$0.047}    &    0.196$\pm$0.175      \\
$\pi_{{\rm E},E}$    &                       &    0.699$\pm$0.031    &    0.921$\pm$0.042    &   {\bf  0.419$\pm$0.031}    &    0.385$\pm$0.029      \\
$\dot{s}$            &                       &                       &                       &   {\bf -0.0651$\pm$0.0020}  &   -0.0693$\pm$0.0022    \\
$\dot{\alpha}$       &                       &                       &                       &   {\bf  0.0064$\pm$0.0068}  &   -0.0004$\pm$0.0101 
\enddata
\tablecomments{
${\rm HJD}'={\rm HJD}-2450000$. 
The parameters of the best-fit model are marked in bold font.
}
\end{deluxetable*}

Characterization of binary lenses requires modeling of lensing light 
curves.  We test 3 different models.  In the first model, we test a 
static binary model (standard model).  In this model, the light curve 
is characterized by 7 parameters.  The first set of three parameters 
are needed to describe the light curves of single-lens events: the 
time required for the source to transit the Einstein radius, $t_{\rm E}$ 
(Einstein time scale), the time of the closest lens-source approach, 
$t_0$, and the lens-source separation in units of the Einstein radius 
at the time of $t_0$, $u_0$ (impact parameter). Another set of three 
parameters are needed to describe the deviation caused by the lens 
binarity: the projected binary separation in units of the Einstein 
radius, $s$, the mass ratio between the binary components, $q$, and 
the angle of the source trajectory with respect to the binary axis, 
$\alpha$ (source trajectory angle). Finally, an additional parameter 
of the ratio of the source radius to the Einstein radius, $\rho_\star
=\theta_\star/\theta_{\rm E}$ (normalized source radius), is needed 
to incorporate the deviation of the light curve caused by the 
finite-source effect.
In the second model, we consider the parallax effect by including two 
additional parallax parameters of $\pi_{{\rm E},N}$ and $\pi_{{\rm E},E}$, 
which are the two components of the microlensing parallax vector 
$\pivec_{\rm E}$ projected on the sky in the direction of north and 
east celestial coordinates.  
In the last model, we additionally check the possibility of the 
effect on the lensing light curve caused by the orbital motion of 
the lens.  The orbital motion affects the lensing magnifications 
in two different ways.  First, it causes the binary axis to rotate 
or, equivalently, makes the source trajectory angle change in time. 
Second, it causes the separation between the lens components to 
change in time.  The latter effect causes alteration of the caustic 
shape in the course of an event.  To first order, the orbital effect 
is parameterized by 
\begin{equation}
\alpha(t)=\alpha(t_0)+\omega \left( {t-t_0\over t_{\rm E}}\right)
\label{eq2}
\end{equation}
and
\begin{equation}
s(t)=s(t_0)+ \dot{s} \left( {t-t_0\over t_{\rm E}}\right)
\label{eq3}
\end{equation}
where the orbital parameters $\omega$ and $\dot{s}$ represent the 
rates of change of the source trajectory angle and the projected 
binary separation, respectively.  Considering the orbital effect 
is important not simply to constrain the orbital motion of 
the lens system but also to precisely determine the lens mass. 
This is because both the motions of the observer (parallax effect) 
and the lens (orbital effect) have a similar effect of causing 
deviations of the source trajectory from a straight line. Then, 
if the orbital motion of the lens is not considered despite 
its non-negligible effect, the deviation of the lensing light 
curve caused by the orbital effect may be explained by the parallax 
effect.  This will cause a wrong determination of the lens parallax 
and the resulting lens mass.

When either the effect of the parallax or orbital motion is considered,
a pair of source trajectories with impact parameters $u_0>0$ and $u_0<0$
result in slightly different light curves due to the breakdown of the 
mirror-image symmetry of the source trajectory with respect to the 
binary axis.  We, therefore, check both models with $u_0>0$ and $u_0<0$ 
whenever the parallax or orbital effect is considered.  As a result, 
the total number of tested models is 5.

\begin{figure}[ht]
\epsscale{1.15}
\plotone{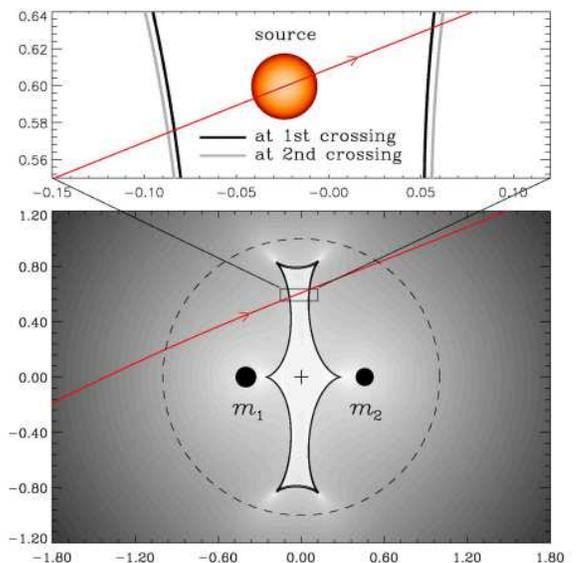}
\caption{\label{fig:two}
Geometry of the binary lens system responsible for the
lensing event OGLE-2005-BLG-153.  In the lower panel, the two
filled dots represent the locations of the binary lens components.
The dashed circle is the Einstein ring corresponding to the total
mass of the binary.  The ring is centered at the center of mass of
the binary (marked by `$+$').  The line with an arrow represents
the source trajectory.  We note that the trajectory is curved due
to the combination of the effects of parallax and lens orbital motion.
The closed figure composed of concave curves represents the positions
of the caustic formed by the binary lens.  All lengths are normalized
by the Einstein radius.  The temperature scale represents the
magnification where brighter tones imply higher magnifications.
The upper panel shows a zoom of the boxed region.  We note that
the caustic shape slightly changes due to the orbital motion of the
lens.  We present the caustics at the two different moments of the
caustic entrance and exit of the source star.
}\end{figure}

To find the best-fit solution of the lensing parameters, we use a
combination of grid and downhill approaches. It is difficult to find
solutions from pure brute-force searches because of the sheer size of
the parameter space. It is also difficult to search for solutions from 
a simple downhill approach because the $\chi^2$ surface is very complex
and thus even if a solution that apparently describes an observed light
curve is found, it is hard to be sure that all possible $\chi^2$ minima 
have been searched. To avoid these difficulties, we use a hybrid approach 
in which a grid search is conducted over the space of a subset 
of parameters (grid parameters) and the remaining parameters are searched 
by a down-hill approach to yield minimum $\chi^2$ at each grid point. 
Then, the best-fit solution is found by comparing the $\chi^2$ values 
of the individual grid points. We set $s$, $q$, and $\alpha$ as grid 
parameters because they are related to the features of light curves 
in a complicated pattern while other parameters are more directly 
related to the identifiable light curve features. For the down-hill 
$\chi^2$ minimization, we use a Markov Chain Monte Carlo method.

\section{Results}

In Table 1, we present the results of modeling along with the best-fit 
parameters for the individual models. It is found that the effects of 
parallax and orbital motion are needed to precisely describe the light 
curve.  We find that the model with the parallax effect improves the 
fit by $\Delta\chi^2=571$.  The fit further improves by $\Delta\chi^2
=164$ with the addition of the orbital effect.  We note that the values 
of the parallax parameters from the ``parallax+orbit'' model are 
different from those determined from the ``parallax'' model. This 
demonstrates that consideration of the orbital effect is important for 
the precise measurement of the lens parallax.

In Figure \ref{fig:one}, we present the model light curve on the top 
of observed data points.  Figure \ref{fig:two} shows the geometry of 
the lens system corresponding to the best-fit solution, i.e.\ 
``parallax+orbit'' model with $u_0>0$.  In the figure, the filled 
dots represent the locations of the binary components, the dashed 
circle is the Einstein ring corresponding to the total mass of the 
lens, the closed figure composed of concave curves is the caustic 
formed by the lens, and the curve with an arrow represents the source 
trajectory. The upper panel shows an enlargement of the region where 
the source trajectory crosses the caustic. The shape of the caustic 
changes in time due to the orbital motion of the lens, and thus we 
present the caustics at two different moments of the caustic entrance 
and exit of the source star.

\begin{figure}[t]
\epsscale{1.15}
\plotone{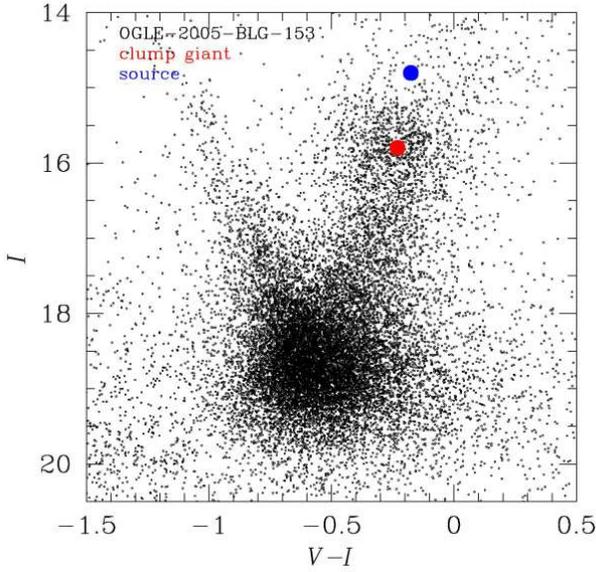}
\caption{\label{fig:three}
Positions of the source (lensed) star with respect to the 
centroid of clump giants in the instrumental color-magnitude 
diagram. 
}\end{figure}

Among the two quantities needed for the determination of the lens 
mass, the microlens parallax is obtained directly from the parallax 
parameters determined from modeling by
\begin{equation}
\pi_{\rm E}=\left( \pi_{{\rm E},N}^2+\pi_{{\rm E},E}^2 \right)^{1/2}
=0.432 \pm 0.042.
\label{eq4}
\end{equation}
On the other hand, the Einstein radius is not directly obtained from 
modeling.  Instead, it is inferred from the normalized source radius 
$\rho_\star$, which is determined from modeling, combined with the 
information about the angular source radius $\theta_\star$.  The 
angular source radius is determined from the information of the 
de-reddened color of the source star measured by using the centroid 
of clump giant stars in the color-magnitude diagram as a reference 
position under the assumption that the source and clump centroid 
experience the same amount  of extinction \citep{yoo04}.  Figure 
\ref{fig:three} shows the instrumental color-magnitude diagram 
constructed based on CTIO $V$ and $I$ band images and the locations 
of the source star and the centroid of clump giants.  By measuring 
the offsets in the color and magnitude between the source and the 
centroid of clump giants combined with the known color and absolute 
magnitude of the clump centroid of $[(V-I),M_I]_{\rm c}=(1.04,-0.25)$, 
we estimate that the de-reddened magnitude and color of the source 
star are $I_0=13.16$ and $(V-I)_0=1.09$, respectively, implying that 
the source is a clump giant with an angular radius of $\theta_\star=
11.72\ \pm 1.01$ $\mu$as.  Here we adopt a Galactocentric distance 
of 8 kpc and the offset of the bar of the field is 0.4 kpc toward 
the Sun and thus the distance to the clump centroid is 7.6 kpc 
based on the Galactic model of \citet{han03}.  Then, with the 
measured normalized source radius of $\rho_\star=0.018\pm 0.001$, 
the Einstein radius is estimated as 
\begin{equation}
\theta_{\rm E}={\theta_\star \over \rho_\star} = 0.66 \pm 0.06\ {\rm mas}.
\label{eq5}
\end{equation}
Together with the Einstein time scale, the relative lens-source 
proper motion is obtained by $\mu=\theta_{\rm E}/t_{\rm E}=5.38
\pm 0.47$ mas ${\rm yr}^{-1}$.

With the measured Einstein radius and lens parallax, the mass of the 
lens system is uniquely determined by 
\begin{equation}
m={\theta_{\rm E} \over \kappa \pi_{\rm E}}=0.19\pm 0.02\ M_\odot,  
\label{eq6}
\end{equation}
where $\kappa=4G/(c^2{\rm AU})$.  With the known mass ratio between 
the binary components, the masses of the individual binary components 
are determined, respectively, as 
\begin{equation}
m_1={1\over 1+q} m=0.10\pm 0.01\ M_\odot 
\label{eq7}
\end{equation}
and 
\begin{equation}
m_2={q\over 1+q}m =0.09\pm 0.01\ M_\odot.
\label{eq8}
\end{equation}
{\it This implies that both lens components are very low mass stars 
with masses just above the hydrogen-burning limit of $0.08\ M_\odot$.}
The distance to the lens is determined as 
\begin{equation}
D_{\rm L}= {{\rm AU}\over \pi_{\rm E} \theta_{\rm E}+\pi_{\rm S}}
=2.42\pm 0.21 \ {\rm kpc}, 
\label{eq9}
\end{equation}
where $\pi_{\rm S}={\rm AU}/D_{\rm S}$ is the parallax of the source 
star.  From this distance to the lens combined with the Einstein radius, 
it is found that the two low-mass binary components are separated with 
a projected separation of 
\begin{equation}
r_\perp=s D_{\rm L} \theta_{\rm E}=1.35 \pm 0.12\ {\rm AU}.  
\label{eq10}
\end{equation}
It is also found that the lens velocity in the frame of the local 
standard of rest is $\textit{\textbf{v}}=(v_\perp, v_\parallel) = 
(-20.9\pm 31.9,15.1\pm 31.9)\ {\rm km}\ {\rm s}^{-1}$, where $v_\perp$ 
and $v_\parallel$ are the velocity components normal to and along the 
Galactic plane, respectively.  We note that the errors in $v$ are 
dominated by the unknown proper motion of the source, which is assumed 
to be $0 \pm 100$ km s$^{-1}$ in the Galactic frame.  The velocity and 
the distance to the lens imply that the lens is in the Galactic disk.

In addition to the parallax effect, the relative lens-source motion 
can, in principle, also be affected by the orbital motion of the 
source star if it is a binary \citep{smith03}.  We check the 
possibility that this so-called ``xallarap'' (reverse of ``parallax'') 
effect influences the parallax determination.  For this, we conduct 
additional modeling including the xallarap effect.  For the description 
of the xallarap effect, 3 additional parameters of the phase angle, 
inclination, and orbital period are needed under the assumption that 
the source moves in a circular orbit.  From this analysis, we find 
that the xallarap effect does not provide a better model than the 
parallax model. In addition, the best-fit occurs for an orbital 
period of $\sim 1$ yr, which corresponds to the orbital period of 
the Earth around the Sun. Furthermore, the best-fit values of the 
inclination and the phase angle are similar to the ecliptic longitude 
and latitude of the source star.  All these facts imply that the 
xallarap interpretation of the light-curve deviation is less likely
and support the parallax interpretation \citep{poindexter05, dong09}.

From the orbital parameters $\omega$ and $\dot{s}$ determined from 
modeling along with the assumption of a circular orbit, one can 
obtain the usual orbital parameters of the semi-major axis, $a$, 
orbital period, $P$, and inclination, $i$, of the orbit of the 
binary lens from the relations
\begin{equation}
a={r_\perp \over x}; \qquad
P=\left( {a^3\over m}\right)^{1/2}; \qquad
\cos i = -(xB)^{1/2}.
\label{eq11}
\end{equation}
Here $B=r_\perp^3 \omega^2/(Gm)$, $x=
\sin\phi$, and $\phi$ is the angle between the vector connecting 
the binary components and the line of sight to the lens such that 
the projected binary separation is $r_\perp=a\sin\phi$.  The value 
of $x$ is obtained by solving the equation $x^3-Bx^2-x+(A^2+1)B=0$,
where $A=(\dot{s}/s)/\omega$ \citep{dong09}.  We find that the 
semi-major axis is $a=1.46\pm 0.08$ AU and the period is $P=4.05\pm 
0.19$ yrs.  The inclination of the orbital plane is $i= 88.3^\circ 
\pm 1.1^\circ$, implying that the orbit is very close to edge on.

We can also constrain the surface brightness profile of the source 
star by analyzing the caustic-crossing parts of the light curve. 
We model the source brightness profile as
\begin{equation}
S_\lambda = {F\over \pi\theta_\star^2}
\left[ 1-\Gamma_\lambda \left(1-{3\over 2}\cos\theta \right) \right],
\label{eq12}
\end{equation}
where $\Gamma_\lambda$ is the linear limb-darkening coefficient and 
$\theta$ is the angle between the normal to the stellar surface and 
the line of sight toward the source star, and $F$ is the source flux. 
We measure the coefficient in $I$ band of $\Gamma_I=0.501\pm 0.014$.
The measured coefficient is consistent with the theoretical value 
of clump giants \citep{claret00}.

\section{Discussion and Conclusion}

We analyze the light curve of a binary-lens microlensing event
OGLE-2005-BLG-153, which exhibits a strong caustic-crossing structure 
on the light curve. By measuring both the Einstein radius and the lens
parallax, we could uniquely measure the masses of the individual lens
components. The measured masses were $0.10\pm 0.01\ M_\odot$ and $0.09
\pm 0.01\ M_\odot$, respectively, and thus the binary was composed of 
very low-mass stars just above the hydrogen-burning limit.

Although the event OGLE-2005-BLG-153 is one of few cases with
well-measured lens masses among the 5000 microlensing events discovered
to date, the event characteristics that enabled this mass measurement
are likely to become common as next-generation microlensing experiments
come on line. Because next-generation experiments will provide intense
coverage from sites on several continents, most caustic-crossing
binaries will yield masses. Moreover, because next-generation cadences
will be independent of human intervention, rigorous characterization of
the selection function will be straightforward. Finally, for reasonable
extrapolations of the mass function of stars close to and below the
hydrogen-burning limit, we can anticipate an important fraction of the 
roughly thousand events per year expected to be detected from 
next-generation surveys to be due to low-mass objects including brown 
dwarfs \citep{gould09}.  Hence, the mass function, at least of objects 
within binaries, will be measurable for all Galactic populations of 
low-mass stellar and substellar objects in the near future, independent 
of biases caused by the luminosity of the population.

\bigskip
We acknowledge the following support: 
National Research Foundation of Korea 2009-0081561 (CH); 
The OGLE project has received funding from the European Research 
Council under the European Community's Seventh Framework Programme
(FP7/2007-2013)/ERC grant agreement number 246678 to AU.
ANR-06-BLAN-0416 (PLANET);
MEXT19015005, JSPS18749004, MEXT14002006, JSPS17340074 (TS); 
Grants MEXT14002006, JSPS17340074, and JSPS19340058 (MOA);
NSF AST-0757888 (AG,SD); 
NASA NNG04GL51G (DD, AG, RP); 
HST-GO-11311 (KS); 
NSF AST-0708890, NASA NNX07AL71G (DPB);
Marsden Fund of NZ (IAB, JBH, DJS, SLS, PCMY); 
Korea Astronomy and Space Science Institute (B-GP, C-UL); 
Dill Faulkes Educational Trust (Faulkes Telescope North); 
The operation of Mt. Canopus Observatory is supported in part by 
the financial contribution from David Warren.
Operation of the Danish 1.54 m telescope at ESO La Silla observatory is
supported by the Danish Natural Science Research Council (FNU)

\end{document}